\documentclass[journal]{IEEEtran}
\usepackage{graphicx}
\usepackage[utf8]{inputenc}
\usepackage[ruled]{algorithm}
\usepackage{algorithmic}
\newlength{\oldparindent}
\newcommand{\myindent}{\par\hspace{\oldparindent}}
\usepackage[flushleft, para]{threeparttable}
\usepackage{multirow}
\usepackage{array}
\newcolumntype{L}[1]{>{\raggedleft\arraybackslash}m{#1}}
\newcolumntype{R}[1]{>{\raggedright\arraybackslash}m{#1}}
\newcolumntype{C}[1]{>{\centering\arraybackslash}m{#1}}
\usepackage{cite}
\usepackage{soul}
\usepackage{color}

\usepackage{float}
\usepackage{amsmath,amssymb,amsfonts,bm}
\allowdisplaybreaks[1]
\usepackage{enumitem}
\usepackage{comment}
\DeclareMathOperator*{\argmin}{arg\,min}

\begin{document}
\bstctlcite{IEEEexample:BSTcontrol}

\title{Towards Expedited Impedance Tuning of a Robotic Prosthesis for Personalized Gait Assistance by Reinforcement Learning Control}

\author{Minhan~Li,
        Yue~Wen,
        Xiang~Gao,
        Jennie~Si,~\IEEEmembership{Fellow,~IEEE,}
        and~He (Helen)~Huang,~\IEEEmembership{Senior~Member,~IEEE
\thanks{This work was supported in part by National Science Foundation \#1563454, \#1563921, \#1808752 and \#1808898. \textit{(Corresponding authors: He (Helen) Huang; Jennie Si.)}}
\thanks{M. Li, Y. Wen,  and H. Huang are with the NCSU/UNC Department of Biomedical Engineering, NC State University, Raleigh, NC, 27695-7115; University of North Carolina at Chapel Hill, Chapel Hill, NC 27599 USA (email: mli37@ncsu.edu; ywen3@ncsu.edu; hhuang11@ncsu.edu).}
\thanks{X. Gao and J. Si  are with the Department of Electrical, Computer, and Energy Engineering, Arizona State University, Tempe, AZ, 85281 USA (email: xgao29@asu.edu; si@asu.edu).}
}
}

\markboth{IEEE TRANSACTIONS ON ROBOTICS}
{Shell \MakeLowercase{\textit{et al.}}: Bare Demo of IEEEtran.cls for IEEE Journals}

\maketitle
\begin{abstract}
Personalizing medical devices such as lower limb wearable robots is challenging. While the initial feasibility of automating the process of knee prosthesis control parameter tuning has been demonstrated in a principled way, the next critical issue is to improve tuning efficiency and speed it up for the human user, in clinic settings, while maintaining human safety. We therefore propose a Policy Iteration with Constraint Embedded (PICE) method as an innovative solution to the problem under the framework of reinforcement learning. Central to PICE is the use of a projected Bellman equation with a constraint of assuring positive semi-definiteness of performance values during policy evaluation. Additionally, we developed both online and offline PICE implementations that provide additional flexibility for the designer to fully utilize measurement data, either from on-policy or off-policy, to further improve PICE tuning efficiency. Our human subject testing showed that the PICE provided effective policies with significantly reduced tuning time. For the first time, we also experimentally evaluated and demonstrated the robustness of the deployed policies by applying them to different tasks and users. Putting it together, our new way of problem solving has been effective as PICE has demonstrated its potential towards truly automating the process of control parameter tuning for robotic knee prosthesis users. 
\end{abstract}


\section{Introduction}
\IEEEPARstart{R}{obotic} prostheses have emerged with recent breakthroughs in mechanical design, control theory and biomechanics \cite{Goldfarb2008,Lenzi2018,HHerr2008,Azimi2019}. These robotic prostheses have manifested exceptional potentials to benefit lower limb amputees in various ways, such as reducing metabolic consumption\cite{HHurr2011}, enhancing balance and stability \cite{Goldfarb2011}, augmenting adaptability to varying walking speeds and inclines\cite{Gregg2018}, and enabling walking on changing terrains seamlessly\cite{TKuiken2013,HHuang2016a,HHuang2013}. The finite-state machine impedance control (FSM-IC) has been the most adopted control framework for prosthetic devices \cite{HHerr2009,Goldfarb2015,Liu2014,Kronander2016}, because studies have suggested that the human nervous system modulates the impedance of lower limb joints in order to realize stable and robust dynamics when walking on various terrains\cite{MKwato2001,Hogan1984}. In addition to the traditional FSM-IC, Azimi et al. recently proposed to estimate the ground reaction force (GRF) \cite{Fakoorian2017}, and integrated it into impedance controller \cite{Azimi2020}, achieving improved tracking performance of a prosthetic knee.  

The challenges in applying FSM-IC in robotic prostheses are: 1) tuning of a large number of impedance control parameters (e.g., 12-15 in a knee prosthesis for level walking only) in order to achieve safe human-machine-environment interaction and sufficient characterization of limb movement within a gait cycle \cite{Goldfarb2008,Winter1991,Liu2014,HHerr2014}; 2) these control parameters must be personalizedto assist individual amputees' gait. To find a modest set of parameters, in current clinical practice, highly trained prosthetists need to spend hours to arduously hand-tune the parameters for each amputee user and each locomotion mode (e.g., level walking, ramp ascent/descent) based mainly on the subjective observations of the user’s gait performance \cite{LHargrove2014}. Not only does it lack precision, but also it needs intensive time and efforts due to the inability of humans to tune high-dimensional parameters simultaneously. 

Given a growing need for facilitating the assistive device personalization, the research community has developed various solutions to automate the process. A few estimation approaches were proposed to determine the impedance parameters by mimicking the nature of biological joints via modeling\cite{Aghasadeghi2013,ERouse2014}. Further, optimization approaches were developed and validated on able-bodied subjects to minimize metabolic cost by using a small number of control parameters for exoskeletons (no more than 4) \cite{JZhang2017,YDing2018}. Beyond merely identifying a set of optimal parameters, a couple of studies have attempted to learn the optimal sequential decision-making for optimizing high-dimensional prosthesis control parameters. Employing knowledge and skills from experienced prosthetists, an expert system was developed to encode human decisions as automatic tuning rules \cite{HHuang2016}. The method was challenged by the lack of sufficient data collected from the prosthetists in device tuning. Alternatively, an actor-critic reinforcement learning (RL) based method (i.e., direct heuristic dynamic programming, dHDP for short) was designed to directly obtain the impedance tuning policy via interaction with the human-prosthesis system in an online manner\cite{YWen2017,YWen2019} without a closed-form model of the system.

Although the aforementioned studies have demonstrated the feasibility of applying automatic tuning to wearable robots with human-in-the-loop, little attention has been paid to address the efficiency and robustness of the tuning algorithms from a user's perspective. First, efficiency of a tuning algorithm (i.e., the ability to safely complete the online tuning rapidly in time) is critical for the clinical translation of a new method due to patient-in-the-loop. Second, the robustness of the tuning algorithm quantifies whether the optimal control parameters or learned prosthesis tuning policy can handle situations when walking condition (e.g., treadmill walking vs. level-ground walking) or user has changed. Additionally, a robust policy is expected to alleviate computational burden in online learning or continued customization, improve user safety during automated prosthesis tuning, and expedite the tuning process in clinics. Therefore, the objective of this study was to develop an efficient and robust automatic tuning method for a robotic prosthetic knee to reproduce near-normal knee kinematics during walking in clinic settings, which include level-ground and ramp. The tuning goal stemmed from the fact that having amputees walk normally as the able-bodied people has been widely used as the design goal or the evaluation criteria for knee prosthesis control\cite{HHerr2014,MVEC2009}. To this end, the objective of this study includes: 1) developing a learning algorithm with enhanced data and time efficiency; 2) investigating the robustness of trained policies against changes in task and user.

To address the efficiency in learning a control parameter tuning policy for robotic prostheses, policy iteration, a classical RL method, lends itself as a promising candidate. This is because, like general RL-based control framework, it has an excellent capability of learning optimal sequential decisions in high-dimensional space\cite{JennieBook}. Also, our approach is data-driven, which learns directly from interactions between the actions and the consequences measured from the human-robot system. In another word, there is no need of explicitly performing a system identification procedure, either online or offline, prior to or during controller design as most control theoretic approaches do, including adaptive control. In addition, as the process of customizing prosthesis control parameter design for a human user does not render abundance of data, which is a necessary feature in those deep RL applications \cite{AZeng2018,Gu2019,Peng2018,Hwangbo2019}, the classic policy iteration framework with moderate demands on data amount is a suitable approach. Furthermore, previous evidences suggest that the policy iteration has the advantage of fast convergence over other classical RL algorithms, such as value iteration and gradient-based policy search\cite{lagoudakis2003,BookFARL} (including our previously reported dHDP \cite{JSi2001} which is a stochastic gradient method). The idea of policy iteration is to iteratively improve the policy by alternately carrying out the policy evaluation and the policy improvement steps \cite{Sutton2018,BertsekasSurvey}. As the efficiency in the policy evaluation step significantly influences the overall learning algorithm efficiency, the problem boils down to improving the efficiency of policy evaluation.

Therefore, in the present study, base upon the policy iteration framework, we proposed an improved solution, namely the policy iteration with constraint embedded (PICE), to enhance policy training efficiency. Such enhancement was enabled by the following special designs in the algorithm tailored for our application. First, we leveraged a projected Bellman equation from which the performance values can be approximately solved during each policy evaluation step. This is to ensure positive semi-definiteness of the value function to avoid incorrect outcome of negative values due to approximation errors. Second, inspired by a widely utilized quadratic cost formulation in successful applications of RL \cite{pieter2006,Lewis2009,Lewis2011}, we adopted simple yet efficient quadratic basis functions to approximately evaluate policies rather than using complex structures such as actor critic networks. Additionally, we provided flexible implementations of PICE to perform either offline or online learning, a proper use of which can further improve learning efficiency. The PICE algorithm was implemented and tested on human-prosthesis systems; the efficiency and robustness of the tuning policy were evaluated quantitatively.

The main contributions of this study are as follows:
\begin{enumerate}
\item We developed a new problem solution PICE based on policy iteration reinforcement learning for improving the efficiency of automatic tuning of the robotic knee control parameters. Our new design entails constraining the performance values from going into the incorrect range of having negative values;
\item The proposed PICE algorithm was implemented and tested in real time on human subjects for prosthesis control parameter tuning. We successfully demonstrated its efficiency and effectiveness in experiments involving human subjects;
\item The robustness of learned control parameter tuning policies against changes of tasks and users were tested on human subjects. The successful demonstration of robustness of PICE suggests its potential values for clinical application.          
\end{enumerate}

The remainder of this paper is organized as follows. Section \ref{PF} describes the problem to be solved and shows how it relates to the theory of optimal sequential decision. Section \ref{RL} presents details of the proposed RL algorithm for improving data and time efficiency. Section \ref{IPLM} elaborates on the considerations regarding the implementations of the algorithm. Results are presented in Section \ref{EXP}. Finally, we discuss these results and limitations of the study in Section \ref{DISS} and conclude in Section \ref{CONCLD}.

\section{Problem Formulation}
\label{PF}
\begin{figure}[tbp]
 	\centering
 	\includegraphics[width=3.45in] {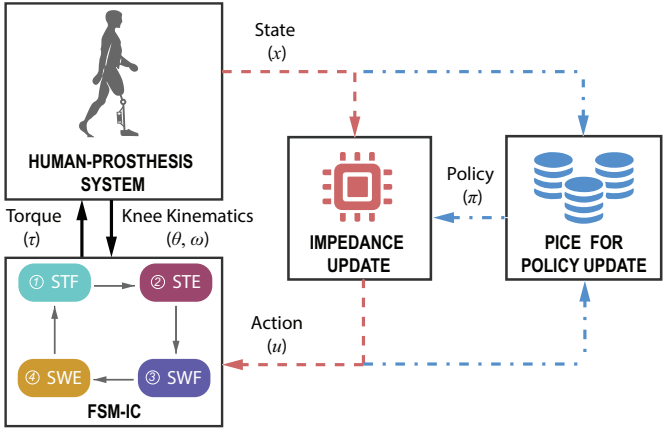}
 	\caption{The schematic illustration of the RL based impedance tuning for a robotic knee prosthesis system within the finite-state machine impedance control (FSM-IC) framework. Red dashed lines denote inputs and outputs in the impedance update loop, whereas blue dash-dotted lines stand for those in the policy update loop. A tuning policy acts to adjust impedance parameters, according to the state of human-prosthesis system, in the FSM-IC to regulate the interaction force with users. The policy can be obtained from and further updated by the proposed PICE algorithm. In the block of FSM-IC, STF, STE, SWF and SWE stand for stance flexion, stance extension, swing flexion and swing extension, respectively.}
 	\label{fig:BlockD}
\end{figure}

In this study, the proposed PICE algorithm aims at determining optimal control parameter tuning policies to supplement the impedance controller of a robotic knee prosthesis in order for its user to restore a near-normal knee motion. The algorithm is implemented within a well-established FSM-IC framework\cite{Goldfarb2008,Liu2014}, as shown in Fig. \ref{fig:BlockD}. 

\subsection{Finite-state Machine Impedance Controller (FSM-IC)}
As depicted in the FSM-IC block of Fig. \ref{fig:BlockD}, a single gait cycle during walking is decomposed into 4 distinct phases in the FSM-IC: stance flexion (STF), stance extension (STE), swing flexion (SWF) and swing extension (SWE). The major gait events determining the phase transitions are identified by utilizing the measurements of knee angle and GRF together using the Dempster-Shafer theory as described in\cite{Liu2014}.

For each phase in a single gait cycle, the FSM selects the corresponding set of impedance parameters for the impedance controller to generate a torque $\tau$ at the prosthetic knee joint based on the impedance control law,
\begin{equation}
\tau=K(\theta_e-\theta)-C\omega
\end{equation}
where the impedance controller consists of three control parameters: the stiffness $K$, the equilibrium angle $\theta_e$ and the damping $C$. Real-time sensor feedback includes the knee joint angle $\theta$ and angular velocity $\omega$. Therefore, a total of 12 impedance parameters need to be regulated in a gait cycle.

\subsection{Dynamic Process of Impedance Update}
As shown in Fig. \ref{fig:BlockD}, the impedance update loop is executed by specified policies to adjust impedance parameters for the FSM-IC. Without loss of generality, the following formulation towards describing the dynamic process of impedance update for a robotic prosthesis is applicable to all four phases in the FSM-IC. This is owing to the fact that, despite sharing the identical framework for learning the tuning policy, each phase is associated with an independent tuning policy running in parallel.  

We consider the human-prosthesis system as a discrete time system with unknown dynamics $f$, which was also studied in \cite{YWen2019,XGao2020}, 
\begin{equation}
\label{system}
\begin{split}
x{\scriptscriptstyle(k+1)} &=f(x{\scriptscriptstyle(k)},u{\scriptscriptstyle(k)}), \; k=0, 1, \ldots\\
u{\scriptscriptstyle(k)} &=\pi(x{\scriptscriptstyle(k)})
\end{split}
\end{equation}
where $k$ denotes the discrete index in the impedance update loop in Fig. \ref{fig:BlockD}. We denote $x$ and $u$ as state and action variables of the process, respectively, while the tuning policy $\pi$ represents a mapping to determine actions according to current states.

\begin{figure}[htbp]
 	\centering
 	\includegraphics[height=2.5in] {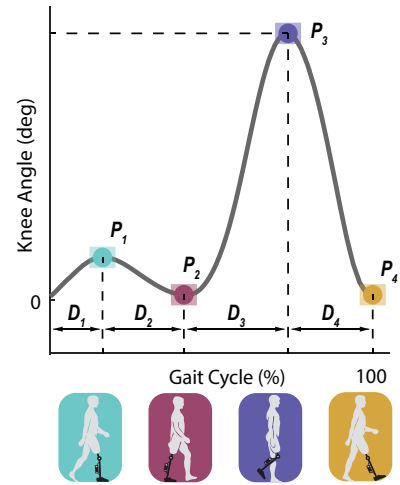}
 	\caption{Features of knee kinematics in a single gait cycle. The subscript numbers 1 through 4 denote respective phases (i.e., STF, STE, SWF and SWE) of a gait cycle, to which the features correspond.}
 	\label{fig:Profile}
\end{figure}

In the context of impedance update, the above state variables are defined based on features extracted from the knee kinematic profiles for each segmented phase in the FSM-IC. Specifically, the continuous knee profile within a single gait cycle (from the heel strike to the next heel strike of the same foot) is characterized by 4 discrete points, each of which is a local extrema in the corresponding phase along the profile as shown in Fig. \ref{fig:Profile}. Each point is associated with two features, the angle feature $P$ and the duration feature $D$, respectively. Similarly, target features ($P^d$ and $D^d$) in each phase can be determined from the representative data of knee kinematics in the able-bodied population\cite{Kadaba1990}.

Consequently, state variables $x\in\mathbb{R}^{2}$ are defined as the differences between measured features and target features (referred as the peak error and the duration error, respectively) in a specific phase at each impedance update as follows,
\begin{equation}\label{rawstate}
x=[P-P^d, D-D^d]^T . 
\end{equation}
Meanwhile, action variables $u\in\mathbb{R}^{3}$ are defined in the following form,
\begin{equation}
\label{rawaction}
u=[\Delta K,\Delta\theta_e,\Delta C]^T ,
\end{equation}
where $\Delta K,\Delta\theta_e,\Delta C$ are the adjustments of impedance parameters for the corresponding phase at each instance of impedance update.

\subsection{Policy Update}
The policy update loop is carried out by the proposed PICE algorithm, as shown in Fig. \ref{fig:BlockD}, to progressively approach optimal policies with respect to specified objectives. In this study, the objective is to regulate states with minimal control energy expenditure over the process of impedance update in order to keep the peak error and duration error as close to zero as possible. 

Hence, at each instance of impedance update, we assign a scalar stage cost with a quadratic form, 
\begin{equation}
\label{cost}
g(x{\scriptscriptstyle(k)},u{\scriptscriptstyle(k)}) =x{\scriptscriptstyle(k)}^{T}R_{s}x{\scriptscriptstyle(k)}+u{\scriptscriptstyle(k)}^{T}R_{a}u{\scriptscriptstyle(k)}
\end{equation} where $R_{s}\in\mathbb{R}^{2\times2}$ and $R_{a}\in\mathbb{R}^{3\times3}$ are both positive semi-definite matrices. Thereby, given a policy $\pi^i$ after the $i$th policy update, the corresponding discounted cost-to-go in an infinite horizon, namely the action-dependent value function $Q^{(i)}$, is given by
\begin{equation}
\begin{split}
Q^{(i)}(x{\scriptscriptstyle(k)},u{\scriptscriptstyle(k)})=&g(x{\scriptscriptstyle(k)},u{\scriptscriptstyle(k)})+\sum_{t=k+1}^{\infty}\alpha^{t-k}g(x{\scriptscriptstyle(t)},u{\scriptscriptstyle(t)}) \\
=&g(x{\scriptscriptstyle(k)},u{\scriptscriptstyle(k)})+\alpha Q^{(i)}(x{\scriptscriptstyle(k+1)},\pi^{i}(x{\scriptscriptstyle(k+1)}))
\end{split}
\end{equation}
where $\alpha$ is the discount factor. This value function is non-negative due to the definition of $g(x{\scriptscriptstyle(k)},u{\scriptscriptstyle(k)})$ in \eqref{cost}, and the value function reflects a measure of the performance when action $u{\scriptscriptstyle(k)}$ is applied at state $x{\scriptscriptstyle(k)}$ and the control policy $\pi^i$ is followed thereafter. 

The goal for PICE, as in value-based RL algorithms\cite{Sutton2018}, is to seek an optimal policy that minimizes the cost-to-go by solving the Bellman optimality equation approximately,
\begin{equation}
\begin{split}
Q^{\ast}(x{\scriptscriptstyle(k)},u{\scriptscriptstyle(k)})&=\min_{\pi}Q(x{\scriptscriptstyle(k)},u{\scriptscriptstyle(k)})\\
&=g(x{\scriptscriptstyle(k)},u{\scriptscriptstyle(k)})+\alpha \min_{u{\scriptscriptstyle(k+1)}}Q^{\ast}(x{\scriptscriptstyle(k+1)},u{\scriptscriptstyle(k+1)})\\
&=g(x{\scriptscriptstyle(k)},u{\scriptscriptstyle(k)})+\alpha Q^{\ast}(x{\scriptscriptstyle(k+1)},\pi^{\ast}({x{\scriptscriptstyle(k+1)}}))
\end{split}
\end{equation}
where $\pi^{\ast}$ and $Q^{\ast}$ denote the optimal tuning policy and the associated optimal value, respectively.

\section{Policy Iteration with Constraint Embedded}
\label{RL}
To solve the above Bellman equation approximately and effectively, we propose the PICE algorithm. Instead of achieving a close approximation to the value function as in most general policy iteration algorithms, the PICE makes use of simple quadratic polynomial basis functions which are expected to provide simplified, albeit with approximation errors, and thus efficient solutions during policy evaluation. Meanwhile, aiming at improving the quality of policy evaluation, the PICE introduces a positive semi-definite (PSD) constraint on the approximated value function to prevent the value function from becoming negative due to approximation errors.

\subsection{Value Function Approximation}

To represent the approximate value function $\hat{Q}^{(i)}$ associated with the policy $\pi^i$ being evaluated (referred to as target policy hereafter), a linear parametric combination of basis functions is often used in classic policy iterations as follows, because of its virtues of easy-to-implement and fairly transparent behavior \cite{lagoudakis2003,BookFARL}.
\begin{equation}
\label{basis}
\hat{Q}^{(i)}(x,u)=\phi(x,u)^{T}r^{(i)}
\end{equation}
where $\phi(x,u)\in\mathbb{R}^{m}$ is the vector of fixed basis functions of states and actions, and the weight parameter vector $r^{(i)}\in\mathbb{R}^{m}$ varies as policy updates. Hereafter, we ignore subscript $k$ in state and action and replace them with $x$ and $u$, respectively, for notation simplicity. Instead of the usual universal approximators, such as multi-layer perceptron neural networks, radial basis functions, and splines, we opt for a simple structure of quadratic polynomials as the basis functions to practically further simplify the basis functions, therefore potentially reducing uncertainties associated with large number of free parameters used in the approximation. 

As a result, the approximating value function can be rewritten in the following equivalent form of weighted inner product, which yields all possible quadratic basis functions of states and actions,
\begin{equation}
\begin{split}
\label{H}
\hat{Q}^{(i)}(x,u)&=
\begin{bmatrix}x\\u
\end{bmatrix}^{T}
H^{(i)}\begin{bmatrix}x\\u
\end{bmatrix}\\
&=\begin{bmatrix}x\\u
\end{bmatrix}^{T}
\begin{bmatrix}H_{xx}^{(i)} & H_{xu}^{(i)}\\
H_{ux}^{(i)} & H_{uu}^{(i)}
\end{bmatrix}
\begin{bmatrix}x\\u\end{bmatrix}
\end{split}
\end{equation}
where $H^{(i)}$ is a PSD matrix, and $H_{xx}^{(i)}$, $H_{xu}^{(i)}$, $H_{ux}^{(i)}$ and $H_{uu}^{(i)}$ are submatrices of $H^{(i)}$ with proper dimensions. By rearranging and grouping like terms in \eqref{H}, we can convert the weight parameter vector $r^{(i)}$ to the matrix $H^{(i)}$ and vice versa. 

\subsection{Policy Iteration Under Constraint}
Two iterative procedures, policy evaluation and policy improvement, are alternately performed in a standard approximate policy iteration. The policy evaluation is to find an approximated value function $\hat{Q}^{(i)}$ satisfying the Bellman equation under the target policy $\pi^i$ as follows,
\begin{equation}
\label{Bellman}
\begin{split}
\hat{Q}^{(i)}(x,u)&=g(x,u)+\alpha \hat{Q}^{(i)}(f(x,u),\pi^i(f(x,u))).
\end{split}
\end{equation}
Replacing the approximate function $\hat{Q}^{(i)}$ with parametric basis functions in \eqref{basis}, we obtain the following equivalent form and its vector-matrix version,
\begin{equation}
\phi(x,u)^{T}r^{(i)}=g(x,u)+\alpha\phi(f(x,u),\pi^i(f(x,u)))^{T}r^{(i)},
\end{equation}
\begin{equation}
\begin{split}
\bm{\Phi}r^{(i)}=\bm{g}+\alpha T\bm{\Phi}r^{(i)}\triangleq B(\bm{\Phi}r^{(i)})
\label{BellmanOp}
\end{split}
\end{equation}
where the matrix $\bm{\Phi}$ consists of basis functions for every possible state-action pair in its rows, and the corresponding stage costs make up the vector $\bm{g}$. In addition, the $T$ and $B$ denote the state transition matrix and the Bellman operator, respectively, under the target policy.

The policy improvement then follows to seek an optimal mapping from states to actions as an improved target policy, with which the next iteration starts,
\begin{equation}
\pi^{i+1}(x)=\argmin_{u\in U}\hat{Q}^{(i)}(x,u)\\
\label{originalImprove}
\end{equation}
where $U$ is the admissible action space. With the choice of quadratic basis functions, the improvement procedure \eqref{originalImprove} is equivalent to solving a quadratic programming (QP) problem. The equivalence can be easily observed by formulating the minimization problem of \eqref{H} over the actions with any given states. As such, the equivalent QP problem can be readily written as
\begin{equation}
\label{policyImp}
\begin{split}
\pi^{i+1}(x)&=\argmin_{u\in U} \{u^TH_{uu}^{(i)}u+2x^TH_{xu}^{(i)}u\} .
\end{split}
\end{equation}

Directly solving the Bellman equation \eqref{BellmanOp} based on the above standard policy iteration framework, we obtained a preliminary proof-of-concept result for offline policy training \cite{MLi2019}. To be more efficient and also to accommodate both offline and online scenarios, we proposed the PICE algorithm with the following details.  

As a result of the approximation error, some of the value functions solved from \eqref{BellmanOp} may yield negative values. This clearly indicates poor approximation given the positive-valued stage cost and the definition of value function. Stemming from insights on the formulated problem, we therefore impose a PSD constraint on the solved value functions from \eqref{BellmanOp} towards an improved solution. Specifically, we seek an approximated value function $\hat{Q}^{(i)}$ satisfying the following projected Bellman equation that is to be solved by PICE,
\begin{equation}
\label{proj}
\bm{\Phi}r^{(i)}=\text{proj}_{S_+}(B(\bm{\Phi}r^{(i)}))
\end{equation}
where $\text{proj}_{S_+}$ denotes the operator of projection onto a closed convex subset $S_+$. The closed convex subset $S_+$ is contained in a subspace spanned by the columns of $\bm{\Phi}$,
\begin{equation}
S_+=\bm{\Phi}R_+
\end{equation}
where $R_+\subset\mathbb{R}^m$ is the PSD cone in the vector space of $\mathbb{R}^m$. 

The idea of solving the projected Bellman equation was also used in the least square policy iteration (LSPI) algorithm \cite{lagoudakis2003}. The PICE algorithm, however, imposes a new and tighter constraint and thus results in a different projected Bellman equation. Specifically, the PICE requires the solved value function from the projected Bellman to be positive semi-defnite. Similar to the LSPI, the PICE can be used as either on-policy or off-policy learning schemes. In general, the behavior policy governing sample distributions for policy evaluation is different from the target policy being evaluated for off-policy learning scheme, whereas they are the same for on-policy learning scheme. Our discussions below on PICE covers both learning schemes. 
 
Inspired by established results \cite{bertsekas2011}, we convert the problem of solving \eqref{proj} to the one that corresponds to the solution of the following variational inequality,
\begin{equation}
(\bm{\Phi}r^{(i)}-B(\bm{\Phi}r^{(i)}))^T\Xi(\bm{\Phi}r-\bm{\Phi}r^{(i)})\geq0, \quad \forall r\in R_+
\label{VI}
\end{equation}
where $\Xi$ is a diagonal matrix and its diagonal elements are the steady-state probabilities of the Markov chain under the behavior policy for each corresponding state-action pair in $\bm{\Phi}$.

For notation simplicity, an equivalent form of the inequality \eqref{VI} can be readily written as
\begin{equation}
\begin{split}
(A^{(i)} r^{(i)}&-b^{(i)})^T(r-r^{(i)})\geq0, \quad \forall r\in R_+\\
& A^{(i)}=\bm{\Phi}^T\Xi(I-\alpha T)\bm{\Phi},\\
& b^{(i)}=\bm{\Phi}^T\Xi\bm{g}.
\label{VI2}
\end{split}
\end{equation}
Involving every single possible state-action pair in terms $A^{(i)}$ and $b^{(i)}$, the inequality is intractable to solve in closed form. Instead, in practice, we can replace the two terms with approximated $\hat{A}^{(i)}$ and $\hat{b}^{(i)}$ by using observational samples as shown in \cite{bertsekas2011},
\begin{equation}
\label{Ab}
\begin{split}
&\hat{A}^{(i)}=\frac{1}{N+1}\sum_{n=0}^{N}\phi(s_n)\Big(\phi(s_n)-\alpha\frac{p_{s_n,s^\prime_{n}}}{q_{s_n,s^\prime_{n}}}\phi(s^\prime_{n})\Big)^T ,\\
&\hat{b}^{(i)}=\frac{1}{N+1}\sum_{n=0}^{N}\frac{p_{s_n,s^\prime_{n}}}{q_{s_n,s^\prime_{n}}}\phi(s_n)g({s_n}) ,\\
\end{split}
\end{equation}
where $n$ and $N$ denote sample index and sample size of the collected data, respectively. The variable $s_n\triangleq(x_n,u_n)$ denotes a sample of a state-action pair, and the variable $s^\prime_n\triangleq(x^\prime_n,u^\prime_n)$ denotes the next sample pair following $s_n$ in the sampling trajectory. In addition, the ratio term $p_{s_n,s^\prime_{n}}/q_{s_n,s^\prime_{n}}$ is in place to correct any mismatch in state transition probability matrix $T$ between the behavior policy and the target policy. Importance sampling can be readily applied to address the mismatch. Specifically, $p_{s_n,s^\prime_{n}}$ and $q_{s_n,s^\prime_{n}}$ denote transition probability from the sample $s_n$ to the sample $s^\prime_{n}$ under the target policy and behavior policy, respectively. In the context of $Q$ value function, the ratio can be further simplified to the following form as shown in \cite{BertsekasSurvey},
\begin{equation}
\begin{split}
\label{IS}
\frac{p_{s_n,s^\prime_{n}}}{q_{s_n,s^\prime_{n}}}=\frac{\delta\big(u^\prime_{n}=\pi^i(x^\prime_{n})\big)}{\nu\big(u^\prime_{n}|x^\prime_{n}\big)}
\end{split}
\end{equation}
where $\delta(\cdot)$ denotes the indicator function (i.e., equals to 1 if $u^\prime=\pi^i(x^\prime)$ and 0 otherwise), and $\nu(\cdot)$ denotes the conditional probability of taking action $u$ of the behavior policy in state $x$.

\subsection{Iterative Approach for Solving Policy Evaluation}
To solve the variational inequality \eqref{VI2}, the following iterative approach has been proposed in previous studies \cite{bertsekas2011,yu2012least} to approximate ${r}^{(i)}$ with $\hat{r}^{(i)}_j$
\begin{equation}
\hat{r}^{(i)}_{j+1}=\text{proj}_{E}[\hat{r}^{(i)}_j-\gamma_j(\hat{A}^{(i)}_j\hat{r}^{(i)}_j-\hat{b}^{(i)}_j)]
\label{Dykstra} ,
\end{equation}
where $j$ denotes iterative steps and $E$ is the constraint set for the solution. 

To result in a convergent sequence, the approach also requires constraint set $E$ to be closed, bounded and convex\cite{yu2012least}. In our case, however, the PSD cone is not bounded. To address this issue, we construct a convex set with an intersection between the PSD cone and an Euclidean ball as follows,
\begin{equation}
E=R_+\cap Z_\delta
\end{equation}
where $Z_{\delta}$ denotes a closed Euclidean ball centered at the origin with the radius of $\delta$, the choice of radius can be as large as needed to cover a sufficient subset of the original PSD cone. Since equation \eqref{Dykstra} involves a projection onto the intersection of two convex sets, the Dykstra's projection algorithm is applied\cite{Dykstra}.

Furthermore, the step size $\gamma_j$ also needs to be decreasing on the order of $1/j$ to guarantee a convergent sequence resulted from \eqref{Dykstra} as follows, 
\begin{equation}
\frac{\gamma_{j}-\gamma_{j+1}}{\gamma_{j}}=\mathcal{O}\Big(\frac{1}{j}\Big) .
\end{equation}

\section{Implementation}
\label{IPLM}
To apply PICE for tuning the impedance control parameters of a prosthetic knee on human subjects, some practical issues need to be considered during implementation. Hereafter, an experimental trial  with a human subject, namely a trial, refers to a single experiment with impedance and policy initializations that allow prosthetic knee control parameters to adapt until reaching a stopping criterion. 
    
\subsection{Human Variability and Stopping Criterion}
\label{StopCrit}
Due to variations in physical conditions and fatigue of human subjects,  measurement noise, and other uncertainties associated with the environment, data recorded from the human-prosthesis system need to be processed and tuning target set needs to be realistically specified. Specifically, impedance update is set to take place every 4 gait cycles to reduce noise introduced by human stride-to-stride variance. That is to say, knee features are averaged over the 4 gait cycles with a single impedance update to form a state-action pair to be used in policy update. In addition, we introduce a target set as tolerance levels of error (specifically, $\pm1.5$ deg for peak errors and $\pm3$ percent for duration errors) to account for the inherent walking variability\cite{Kadaba1990,WINTER1984}. Consequently, we consider an impedance parameter tuning procedure in a given phase a success if the errors stay within the target set for 8 out of 10 consecutive impedance updates. If all four phases become successful, a trial is successful and is considered reaching the stopping criterion.  

\subsection{Safety Bounds for Impedance Tuning}
In each trial, a set of initial impedance parameters are selected for the prosthetic knee, and then subjects experience a series of impedance updates guided by tuning policies for each phase of the FSM. While the initial impedance parameters are randomly selected, they need to be feasible for walking. Such feasible initial impedance parameter setting is validated prior to the start of a trial and is verified by the experimenter either via visual inspection if the subject is capable of walking without holding on a handrail, or via the subject’s verbal expression. To avoid any potential harm to human subjects caused by unsafe parameters and associated knee kinematics, we set a safety range within which peak error is allowed to vary. Once a peak error is beyond the safety range, impedance parameters will be reset to the initial ones, which are known to be safe. Herein, the peak error bounds are set to $\pm12$ deg for all four phases since they cover two standard deviations of knee kinematic features in normal walking among different test subjects\cite{Kadaba1990}. More importantly, the safety range also defines a compact set for states and actions, which consequently guarantees our implementation to fulfill the requirement of initial admissible policy for general policy iteration algorithms because the zero-output policy is always an eligible initial policy.  

\subsection{Implementations of PICE}
Prior to feeding data into the PICE algorithm for policy training, feature scaling was first performed on state and action variables for all four phases. To normalize them into a comparable unit magnitude, following scaling factors were selected in the study. Specifically, the state variables $x$ in \eqref{rawstate} were normalized with a scaling factor of $8$ and $0.24$ for the respective peak error and duration error. Similarly, the action variables $u$ in \eqref{rawaction} were normalized with a scaling factor of $0.05$, $0.5$ and $0.0005$ for respective adjustments of stiffness, equilibrium angle and damping. The only one exception was that the value for equilibrium angle in the SWF was set to $1$ when considering phase differences. Meanwhile, to keep actions staying in a reasonable range, we set the admissible space $U$ in \eqref{policyImp} for normalized action variables to a range between $-1$ and $1$.

The PICE features a flexibility that can be implemented in both offline and online manners by taking off-policy and on-policy learning schemes, respectively. The procedures of both implementations are described as pseudocodes in Algorithm \ref{offalg} and Algorithm \ref{onalg}. A summary of value selections for parameters in PICE implementations is listed as follows. Penalty matrices for states $R_s$ and actions $R_a$ were set to $diag(1, 0.5)$ and $diag(0.01, 0.01, 0.01)$, respectively, while discount factor $\alpha$ was selected as $0.9$. The radius of Euclidean ball $\delta$ was assigned to $100$. The tolerance for offline training $\varepsilon_{a}$ was set to $10^{-4}$. The batch size $N_b$ for online training samples was selected as 15. 

\begin{algorithm}[tbhp]
\caption{Offline Off-Policy PICE} 
\label{offalg}

\par\textbf{Initialization: }
\myindent Random initial target policy $\pi^{0}$, policy update index $i\gets 0$; 
\myindent Empty replay buffer \textit{DS} with capacity $N$;
\myindent Choose a tolerance for offline training $\varepsilon_{a}$.

\par\textbf{Offline Data Preparation: }
\myindent Populate the buffer \textit{DS} with $N$ samples of 4-tuple $(x_n,u_n,g_n,x^{\prime}_n)$ generated by a behavior policy from previous experiments, where $n$ is sample index and $x^{\prime}$ denotes the next state in each sample. 
\par\textbf{Iteration:}
\begin{algorithmic}[1]
\REPEAT
\STATE \textbf{(Update Next Actions)} Compute $u^{\prime}_n$ with current target policy $\pi^i(x^{\prime}_n)$ for every sample in \textit{DS};
\STATE \textbf{(Update Sampling Weight)} Compute the ratio $\frac{p_{s_n,s^\prime_{n}}}{q_{s_n,s^\prime_{n}}}$ by \eqref{IS} for every sample in \textit{DS};
\STATE \textbf{(Policy Evaluation)} Evaluate policy $\pi^i$ by solving \eqref{Dykstra} for $\hat{r}^{(i)}$ and $\hat{Q}^{(i)}$, along with approximating \eqref{Ab} using all samples in \textit{DS};
\STATE \textbf{(Policy Improvement)} Update policy $\pi^{i+1}$ by \eqref{policyImp}, $i\gets i+1$;
\UNTIL{$\lVert\hat{r}^{(i)}-\hat{r}^{(i-1)}\rVert_2\leqslant\varepsilon_{a}$};
\RETURN $\hat{Q}^{\ast}\gets\hat{Q}^{(i)}$ and $\hat{\pi}^{\ast}\gets\pi^{i}$.
\end{algorithmic}
\end{algorithm}

\begin{algorithm}[tbhp]
\caption{Online On-Policy PICE} 
\label{onalg}
\par\textbf{Initialization:}
\myindent Choose a batch size $N_b$, and empty replay buffer \textit{DS};
\myindent Random initial target policy $\pi^{0}$, policy update index $i\gets 0$;
\myindent Random initial state $x_{(0)}$, impedance update index $k\gets 0$;
\myindent Choose an initial action $u_{(0)}$ using policy $\pi^{0}(x_{(0)})$;
\par\textbf{Iteration: }
\begin{algorithmic}[1]
\REPEAT
\STATE \textbf{(Online Data Collection)}
Execute the action $u_{(k)}$, observe stage cost $g_{(k)}$ and next state $x_{(k+1)}$, choose the next action $u_{(k+1)}$ by following current target policy $\pi^{i}$, form a 5-tuple sample $(x_{(k)},u_{(k)},g_{(k)},x_{(k+1)},u_{(k+1)})$ and store it in \textit{DS};

\IF{$k=iN_b (i\in\mathbb{N})$}
\STATE \textbf{(Policy Evaluation)} Evaluate policy $\pi^i$ by solving \eqref{Dykstra} for $\hat{r}^{(i)}$ and $\hat{Q}^{(i)}$, along with approximating \eqref{Ab} using all samples in \textit{DS};
\STATE \textbf{(Policy Improvement)} Update policy $\pi^{i+1}$ by \eqref{policyImp}, $i\gets i+1$;
\STATE \textbf{(Reset Buffer)} Empty \textit{DS};
\ENDIF
\STATE $k\gets k+1$;
\UNTIL {Early-stopping  termination  condition is fulfilled}
\RETURN $\hat{Q}^{\ast}\gets\hat{Q}^{(i)}$ and $\hat{\pi}^{\ast}\gets\pi^{i}$.
\end{algorithmic}
\end{algorithm}

The two implementations differ from each other in two major aspects: training data and termination condition. For offline PICE, the training data consist of a fixed set of 4-tuple samples, which were collected from previous studies and already stored in a buffer beforehand. The same data are repeatedly utilized until the iteration terminates. The off-policy scheme is adopted for offline training as the original behavior policy generating the training samples is different from the target policy to be evaluated for every iteration. On the other hand, the online PICE follows an on-policy scheme. In each online iteration, the target policy keeps interacting with the human-prosthesis system to generate on-policy samples of up the specified batch size, thereby performing the policy evaluation correspondingly. The sample will be discarded after usage and replaced with new ones for the new target policy.

In addition, in contrast to offline PICE which terminates iterations until the parameters in the parameterized value function get converged, we used an early-stopping termination condition to deactivate an online training process. It is not only to prevent over training, but also to take into consideration that human subjects can only walk for about 30-60 minutes during an experiential trial due to physical constraints. Specifically, during online training, we analyze the trend in evolution of the stage cost based on the current policy every time when we have newly collected $N_b$ samples of state-action pairs. When either of the two conditions listed below is fulfilled, the online training is deactivated and the rest of impedance update is carried out with the current policy until policy update is triggered again.
 
\begin{enumerate}
\item The case of no occurrence of impedance reset due to hitting the safety bounds. We fit a linear regression model between the time series of stage cost and that of impedance update. From the model, we obtain a confidence interval (specifically $95\%$ confidence interval) around the slope of the regression line. If the interval falls below zero, which signals a rigorously decreasing stage cost as impedance parameters updated according to the current policy, we deactivate the online training;
\item The case of using stage costs. We averaged the stage costs over samples being analyzed. If it is smaller than a threshold value $\varepsilon_{b}$, online training is deactivated. We set the threshold to 0.043, which is equivalent to the stage cost of the largest tolerated errors within target set (i.e., $1.5$ deg for peak error and $3$ percent for duration error).
\end{enumerate}

\section{Experiments and Results}
\label{EXP}
We performed three tests involving four human subjects (two able-bodied and two amputees) to evaluate the performances of the proposed PICE.
\subsection{Hardware Setup}
\begin{figure}[tbph]
 	\centering
 	\includegraphics[width=1.5in] {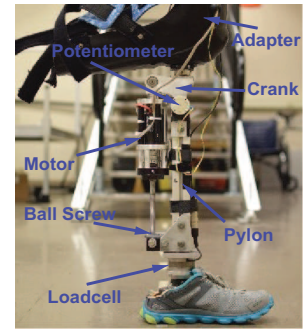}
 	\caption{Hardware setup for the prototype of robotic knee prosthesis.}
 	\label{fig:Hardware}
\end{figure}

A prototype of robotic knee prosthesis designed in our lab was used in this study\cite{Liu2014}. The prosthesis utilizes a slider-crank mechanism, in which the slider is driven by the rotation of a DC motor (Maxon, Switzerland) through a ball screw (THK, Japan), and the crank rotation mimics the knee motion. The whole mechanism is integrated with a pylon as shown in Fig. \ref{fig:Hardware}. A maximum of 80 Nm torque output at the joint is ensured with such a design. The rotational motion of the prosthetic knee joint is recorded by a potentiometer (ALPS, Japan). A load cell (Bertec, USA) is attached to the pylon to measure the GRF. All the analog readings are converted to digital signals through a DAQ board (NI, USA) and then fed back to the control system, which is implemented by LabVIEW and MATLAB on a desktop PC.

\subsection{Participants}
We recruited four male subjects, two able-bodied individuals (AB1 and AB2) and two transfemoral amputees (TF1 and TF2), in this study. The participants' information is summarized in Table \ref{Tab1}. An L-shaped adapter (see Fig. \ref{fig:Hardware}) and a daily socket were used by AB and TF subjects, respectively to allow them to walk with the knee prosthesis. The alignment of prosthesis for each subject was done by a certified prosthetist. All the subjects received training with the powered prosthesis until they can walk comfortably and confidently without holding the handrail. All the subjects were provided written informed consent before any procedures, and the study was approved by the Institutional Review Board of University of North Carolina at Chapel Hill.  

\renewcommand{\arraystretch}{1.2}
\begin{table}[!bph]
\centering
\begin{threeparttable}
\caption{Participant Information}
\begin{tabular}{C{1cm} C{0.5cm} C{1.2cm} C{1.2cm} C{1.2cm}}
\hline \hline
Subject  & Age & Height &  Body Weight$\ast$ & Prosthetic Side \\ 
\hline
AB1 & 43 & $1.78$ m & $74$ kg & Right \\ 
AB2 & 32 & $1.78$ m & $65$ kg & Left  \\ 
TF1 & 63 & $1.64$ m & $64$ kg & Left  \\ 
TF2 & 59 & $1.83$ m & $94$ kg & Left\\ 
\hline \hline
\end{tabular}
\label{Tab1}
\begin{tablenotes}[flushleft]
\item$\ast$ The body weight includes the robotic knee prosthesis.
\end{tablenotes}
\end{threeparttable}
\end{table}

\subsection{Experimental Protocols}
We carried out three experimental tests to validate and analyze the performance of the proposed PICE. They are respectively associated with the following three goals: 1) to experimentally assess convergence properties during offline training and the effect of training data size; 2) to quantitatively assess potential gains by using an offline pre-trained policy as the initial policy for online training, and compare its performance to randomly initialized online training; 3) to investigate the robustness of a set of well-trained policies as tasks and users change. 

\subsubsection{Test of Offline Training}
We used five sets of offline data all collected from AB1 to perform the offline policy training and obtained five sets of policies accordingly. The numbers of data samples in the five sets were 15, 45, 75, 105 and 135, respectively, each sample was a 4-tuple $(x_n,u_n,g_n,x^{\prime}_n)$. We then evaluated each policy in five independent trials using AB1 as the test subject. AB1 walked on a treadmill at a speed of 0.6 m/s, while the offline trained policy adjusted the prosthesis impedance. The same set of initial impedance was applied to all the five trials. To eliminate the confounding effect of fatigue resulting from prolonged walking, for each tuning trial, AB1 performed several 3-minute walking segments followed by a rest period. Additionally, a maximum of 135 impedance updates were allowed in consideration of the subject's limited enduring with walking. If training did not complete within this limit, the trial was considered a failure.

Two outcome measures were captured in each trial. The first measure was the $L^{2}$ distance (i.e., $\lVert{r}^{(i)}-{r}^{(i-1)}\rVert_2$) between the two consecutive weight parameter vectors in \eqref{VI2}. It is used to quantify changes in the series of value outcomes in \eqref{Bellman}. The second measure was to explore the relationship between the number of offline training samples and the number of phases in which the success as defined in Subsection \ref{StopCrit} was reached without any online policy updates beyond offline training.

\subsubsection{Test of Online Training} 
We conducted online training under two different initial policy conditions: 1) randomly initialized; 2) offline pre-trained. Two subjects, AB1 and TF1, were asked to perform the treadmill walking task at the speed of 0.6 m/s. We used their own available offline data to obtain the pre-trained policies. For both subjects, the offline training data had 105 samples. The same pre-trained policies were used to serve as initial policies across trials for each subject, while randomly initialized policies varied. A few blocks of experimental sessions were conducted, each including two online training trials for comparison purpose. Specifically, in each block, we randomly selected the initial impedance parameters with the only requirement of being feasible for walking. Then two online training trials with different initial policy conditions were performed. For AB1, three blocks of experimental sessions (each of which used different initial impedance parameter) were conducted. For TF1, one block was tested. The same walk-rest experimental protocol and the same maximum number of impedance update were applied as discussed in the first test.

The evaluation for the test of online training included efficiency, effectiveness, and impedance tuning convergence. Herein, the efficiency of online training was quantified by: 1) the number of phases needed for online policy updates beyond the initial policies until meeting the stopping criteria defined in Subsection \ref{StopCrit}; 2) the number of impedance updates to meet the stopping criterion for prosthesis tuning. To understand the effectiveness of tuning prosthesis control for producing desired knee motion, the knee kinematics were measured to reflect how the prosthetic knee joint moved when it interacted with the human users as the impedance varied with the guidance of policies. Finally, the impedance tuning convergence was analyzed by checking the evolution of peak errors and duration errors of knee kinematics (states) and prosthesis impedance values (control parameters) during the tuning.

\subsubsection{Test of Policy Robustness} 
We investigated the robustness of well-trained policies and studied how well they behaved against changes of task and human subject. Two sets of well-trained policies were used, which were obtained from AB1 and AB2 in their respective treadmill walking tasks at the speed of 0.6 m/s prior to trials in this test. We applied them as initial policies for three new trials. Specifically, the trials consisted of: 1) up-ramp walking with slope of 4 degrees performed by subject AB1 starting with his own policy; 2) self-paced level-ground walking performed by subject AB2 starting with his own policy; 3) treadmill walking (0.6 m/s) performed by subject TF2 but starting with AB2's policy. To investigate how policies acted when they were applied to different tasks or subjects, we monitored sequences of both impedance and policy updates in each trial and the associated evolution of stage cost.  

\subsection{Experimental Results}

\begin{figure}[!tbhp]
 	\centering
 	\includegraphics[width=\columnwidth] {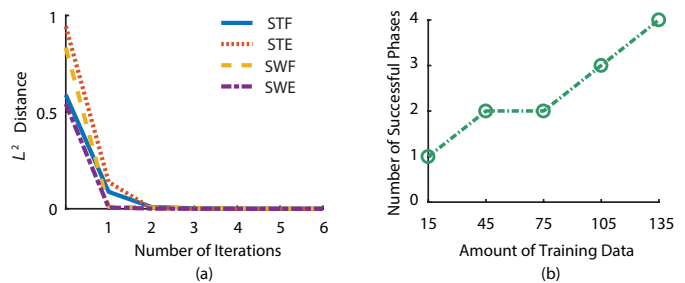}
 	\caption{Result of offline training evaluation. (a) Evolution of the $L^{2}$ distance between weight parameter vectors in two value function consecutive updates (i.e., $\lVert{r}^{(i)}-{r}^{(i-1)}\rVert_2$) in a representative trial of offline training test. At each instance of offline policy update, the vector is updated correspondingly. (b) The number of phases being able to reach a success increases with the increase of amount of training data.}
 	\label{fig:Offtraining}
\end{figure}

\subsubsection{Offline Training Assessment}
Since similar results were obtained from all trials in the test of offline training, we only demonstrate the representative results (trained by data with the size of 105 samples collected from AB1) in Fig. \ref{fig:Offtraining}(a). As shown in the data that changes in the weight parameter vectors of the approximating value function in \eqref{VI2} reduced to within the tolerance (10$^{-4}$) in 6 offline policy updates, which is equivalent to 10 seconds for performing the computation. The results suggest that the approximate value functions, as well as the policies, were convergent given the offline training data.

For the effect of the size of training dataset, we see in Fig. \ref{fig:Offtraining}(b) that the number of phases being able to reach a success increased as the amount of training data increased. Particularly, when we performed the offline training with 135 samples, the number of phases reached up to four and no more policy updates were needed to accomplish the tuning. The evidence implies that, with the offline implementation alone, the proposed PICE algorithm is able to obtain policies ready to deploy as sufficient offline data of good quality are available to use.

\subsubsection{Online Training Assessment} 
We first looked into its improvement in tuning efficiency by employing the pre-trained initial policies obtained from offline training. As shown in Table \ref{Tab2}, the comparison results reveal that online training starting with pre-trained policies obtained from offline training, albeit not perfect, were significantly more efficient than those starting with random policies. On average, the former cases only resulted in 1 phase that required online policy update, whereas the number amounted to 4 in the latter cases. Meanwhile, pre-trained cases were observed to have less overall number of impedance updates than random cases did to meet the stopping criterion by an average of 58, which was equivalent to about 7 minutes of walking time of a subject.  

\newcommand{\minitab}[2][@{}c@{}]{\begin{tabular}{#1}#2\end{tabular}}
\newcommand\mypound{\scalebox{0.9}{\#\hspace{0.1cm}}}
\begin{table}[!tb]
\centering
\begin{threeparttable}
\caption{Efficiency Comparisons of Online Training by Using Pre-trained vs. Random Initial Policies}
\label{Tab2}
\begin{tabular}{cccccc}
\hline\hline
\multirow{2}{*}{\minitab{Experimental\\Block Number\tnote{$\ast$}}} & \multicolumn{2}{c}{\minitab{Phase Requiring\\Policy Updates\tnote{$\dagger$}}} & & \multicolumn{2}{c}{\minitab{Overall Number of \\Impedance Updates}}\\ 
\cline{2-3} \cline{5-6}
& Pre-trained  & Random &  & Pre-trained  & Random \\ \hline
\centering B1 & 2 & 1, 2, 3, 4 &  &39 & 126\\  
\centering B2 & 2 & 1, 2, 3, 4 & & 43 & 94\\ 
\centering B3 & 2 & 1, 2, 3, 4 &  &42 & 111\\ 
\centering B4 & 4 & 1, 2, 3, 4  & & 28 & 53\\ 
\hline\hline
 \end{tabular}
\begin{tablenotes}
\item\text{$\ast$} The experimental block B1 through B3 were tested with subject AB1,\\ \hspace*{2.2mm} while the block B4 was associated with the subject TF1.\\
\item\text{$\dagger$} The numbers 1 through 4 under the column represent STF, STE, SWF \\ \hspace*{2.2mm} and SWE respectively in the FSM-IC.
\end{tablenotes}
\end{threeparttable}
\end{table}

\begin{figure}[!tbhp]
 	\centering
 	\includegraphics[width=\columnwidth] {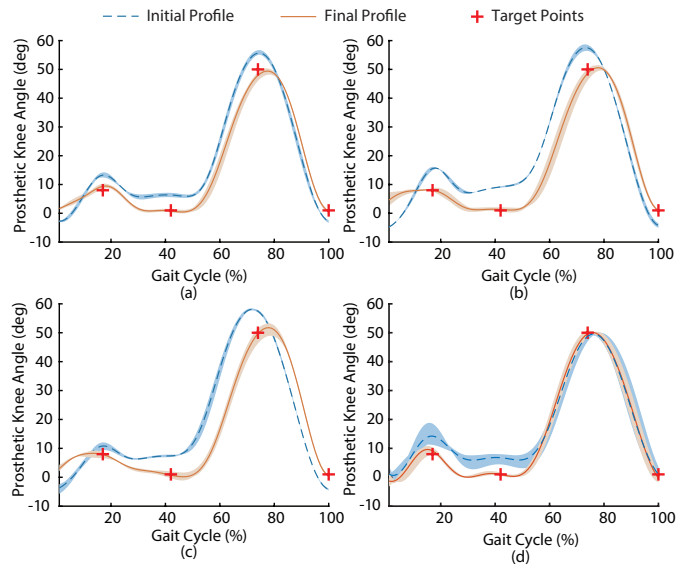}
 	\caption{Prosthetic knee kinematics with initial and final tuned impedance parameters in the test of online training. (a) to (d) Trials with pre-trained initial policies in experimental block B1, B2, B3 and B4, respectively. Time series of kinematics are divided and normalized to multiple profiles in individual gait cycles based on the timing of heel strike. Shaded areas along profiles indicate the real motion ranges across 4 gait cycles performed by subjects walking with the same impedance parameters. The associated lines (dashed and solid) denote the averaged kinematics.}
 	\label{fig:BA}
\end{figure}

Apart from the efficiency, for the trials staring with pre-trained initial policies, we studied the tuning effectiveness. Fig. \ref{fig:BA} displays the overall effect of tuning by comparing the knee profiles generated by initial impedance parameters before tuning with those produced by adjusted parameters at the end of tuning trials. We noted that, though differed in shape, initial knee profiles in all trials deviated from the targets, especially peak angle features. However, going through the tuning process under the guidance of final policies, the final parameters enabled knee profiles to approach the targets.

\begin{figure}[!tbhp]
 	\centering
 	\includegraphics[width=\columnwidth] {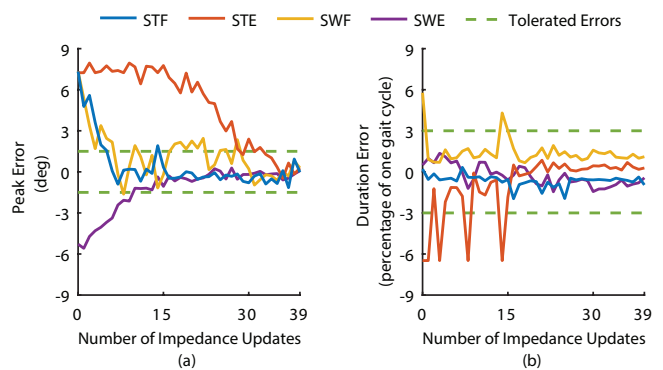}
 	\caption{Evolution of states as impedance parameters were updated. (a) Peak errors, (b) Duration errors.}
 	\label{fig:State}
\end{figure}

\begin{figure*}[!tbhp]
 	\centering
 	\includegraphics[width=1.9\columnwidth] {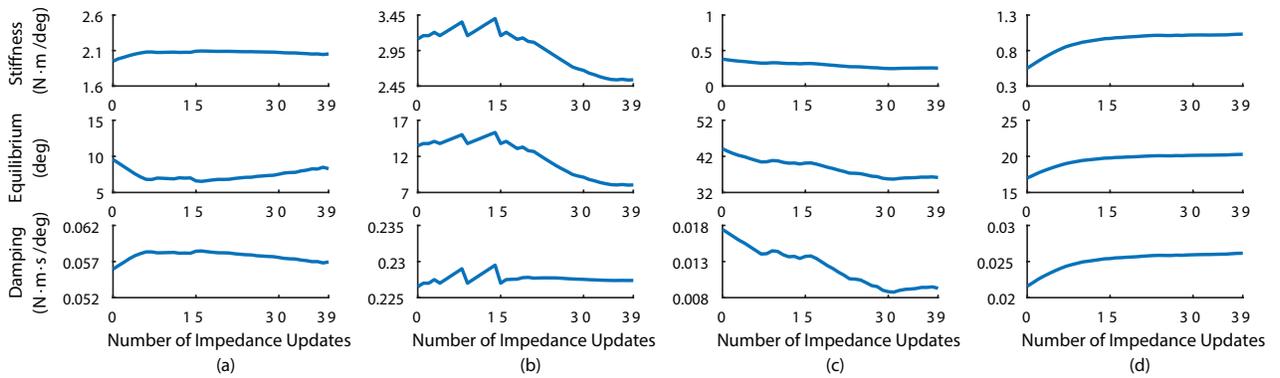}
 	\caption{Evolution of the impedance parameters (i.e., stiffness, equilibrium angle and damping) in different phases. (a) STF, (b) STE, (c) SWF, (d) SWE.}
 	\label{fig:IP}
\end{figure*}

To inspect the impedance tuning convergence, we present representative results here (the experimental block B1 with pre-trained initial policies in the Table \ref{Tab2}) as similar results across trials were observed. As revealed in Fig. \ref{fig:State}, no matter how large the initial errors were, they all progressively converged into the tolerance range of errors ($\pm$1.5 degrees for peak error, $\pm$3 percent for duration error) and eventually remained within the range. Correspondingly, in Fig. \ref{fig:IP}, we observed that impedance parameters converged to constant values at the end of the trial (i.e., last ten updates) for most phases, except for the STF where the momentum of impedance adjustment lingered. The difference may be attributed to varying perturbations introduced by more dynamical interactions occurring in the STF among the human, the robotic prosthesis and the ground. As a result, the final policy for STF needed to respond by adjusting the impedance to accommodate such disturbances and stabilize errors within tolerances.   

\subsubsection{Robustness Investigation}
\begin{figure*}[tbph]
 	\centering
 	\includegraphics[width=2\columnwidth] {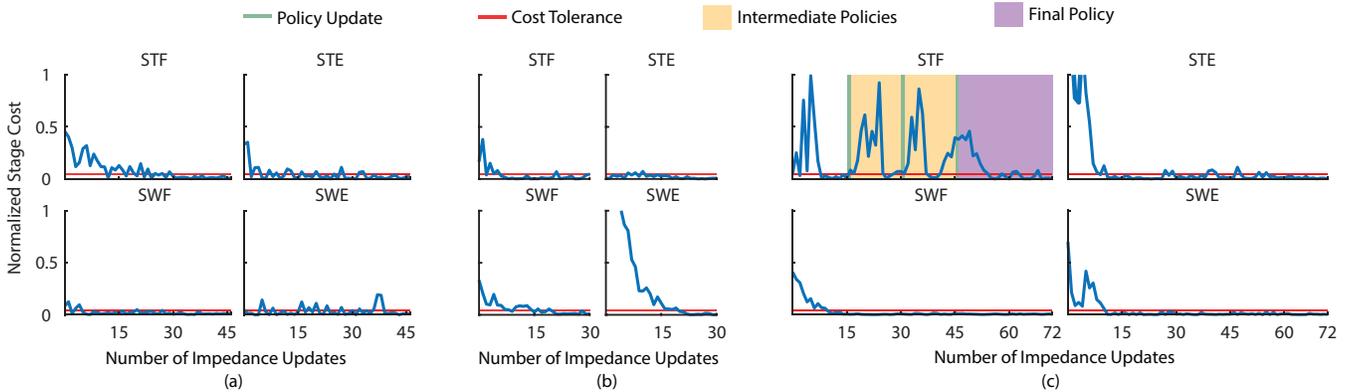}
 	\caption{Normalized stage cost over the number of impedance updates in testing policy robustness. The vertical lines indicate the instances where policy updates took place. The areas highlighted in yellow denote the periods of time when the online learning was activated and intermediate policies were employed, the areas highlighted in purple describe the phases where online learning was deactivated and final policies were deployed. Remaining areas without highlights indicate that initial policies trained from the treadmill walking of one subject were sufficient for successful impedance tuning when applied directly to other walking tasks or subjects. (a) Four phases in a trial of AB2 level-ground walking, (b) Four phases in a trial of AB1 up-ramp walking with slope of 4 degrees, (c) Four phases in a trial of TF2 treadmill walking.}
 	\label{fig:Robust}
\end{figure*}

As seen in Fig. \ref{fig:Robust}(a), subject AB2 used a pre-trained policy obtained from his own treadmill walking to perform level-ground walking with no difficulty as no further policy update was needed, and after 46 impedance updates (about 6 minutes of subject walking time), the subject knee kinematics met stopping criterion. Similar results were observed from Fig. \ref{fig:Robust}(b) in the trial of AB1 up-ramp walking with even fewer impedance updates (i.e., 30 impedance updates). As for the trial of TF2 treadmill walking using the AB2's pre-trained policy, despite not being completely successful in deploying policies to all four phases, only three updates of policy occurred in the STF phase, as shown in Fig. \ref{fig:Robust}(c). Although 72 impedance updates (about 9 minutes) were needed to meet the stopping criterion, it only took 45 updates of impedance (about 6 minutes) to obtain the final policy refined for the STF phase of the new subject. 

Note that a cyclic pattern of change in the cost was displayed in Fig. \ref{fig:Robust}(c). This was caused by following initial or intermediate policies, which led to cost value sloping upward until the safety bound was hit, thereby triggering the reset of impedance parameters and getting the cost drop back to the initial value. The results suggest that policies we obtained from AB2 treadmill walking were, to some extent, robust against the changes of tasks and subjects. 

\section{Discussions}
\label{DISS}
In this study, we proposed a promising solution of PICE for tuning high-dimensional robotic knee prosthesis control parameters in order to provide efficient personalized assistance in walking. The tuning efficiency stemmed partly from our innovation that enables offline policy training, beside online training, via policy iteration. To our knowledge, few studies have successfully addressed wearable robot personalization in such an offline-online manner. 

Our proposed RL method, as suggested in Fig. \ref{fig:Offtraining}, demonstrated the feasibility of obtaining policies from a sufficient amount of existing offline data by the offline training, which can be deployed directly without interacting with the real human-prosthesis system. Clearly the offline implementation of PICE enables a maximal utility of existing parameter-performance data and is a new way to improve training efficiency in obtaining prosthesis tuning policies. Nevertheless, pure offline implementation has no guarantees to obtain accurate and robust policies despite being convergent in the sense of offline training, especially when the training data quality is poor. Note that the quality of data has two meanings, which include the amount of data and the extent of mismatch in data distribution\cite{YLiu2018}. In this study, however, we only investigated the influence of the amount of data on the offline training. Therefore, the number of training data we examined in this study might not be applicable to other datasets due to the confounding effect caused by data distributions, and it is actually difficult to determine the exact number in practice. Hence, an RL algorithm that is capable of performing offline-online training, such as our proposed PICE, became especially intriguing in order to ensure efficiency and effectiveness of auto-tuning algorithm for learning the prosthesis tuning policy. In this paper, we demonstrated that when offline learned policies cannot handle realistic human-prosthesis interaction or were not robust enough to handle the variation across human users, as shown in Fig. \ref{fig:Robust}(c), PICE can trigger online training that further update the policy to achieve the desired tuning goal. 

In addition, the investigation of robustness associated with policies learned by the proposed algorithm showed other indirect benefits to potentially scale up the training outcome. As shown in Fig. \ref{fig:Robust}, most deployed polices possessed exceptional robustness, in spite of the fact that policy refinements happened in the STF through online training to further accommodate for the changes in users. A potential reason to explain the phenomenon could be the fact that the underlying physical principles in prosthesis control have no drastic changes across different subjects and walking tasks as well as the associated variations in gait patterns and GRFs. The promising discovery may enable us to collect data and obtain pre-trained initial policies, albeit not optimal, from more available users and relatively easier tasks. From there, further user-specific or task-specific refinements, if needed, could be accomplished by online training. As opposed to learning from scratch, such an approach is more likely to result in higher training efficiency, and thus it is of great clinical value when applied at scales.

As a generic and efficient learning framework, the proposed PICE could also potentially shed light on similar problems for other assistive wearable machines, such as exoskeletons, neuroprosthetics. These devices are also in need of identifying the optimal control parameters for individual users with motor deficits\cite{O'Malley2016,FES2011}. By unleashing the potentials demonstrated in this study, translations of the proposed approach into other human-machine systems are expected to be valuable because they all call for high training efficiency and being model-free due to patient-in-the-loop. However, specific modifications regarding problem formulations or implementations need to be properly considered before the translations, such as how to define states, actions and costs for each application. 

The successful implementation of the proposed PICE in the human-prosthesis system would encourage future studies to explore more application-specific solutions to an efficient approximation of the value function in reinforcement learning. We demonstrated, in the study, that leveraging simple basis functions (e.g., quadratic basis functions) fueled by insights on the control problem (e.g., the PSD constraint for the value function presented in this paper) is likely to yield a satisfying approximation for the value function with limited amount of data. This is because, with fewer number of unknown parameters to estimate, such a choice alleviates the high demands for persistent excitation\cite{FrankL2010} or data richness\cite{Modares2014} required by generic basis functions, which are often difficult to meet in practice; meanwhile the pre-structured treatment is able to compensate for a poor approximation due to the lack of data.     

Our proposed design and study, although promising, also had several limitations. The primary limitations of the study were the limited evaluation of the algorithm on human subjects and walking terrains in daily environments, because the focuses of the study were on developing a new automatic prosthesis tuning solution and demonstrating its promising advantages in clinic settings. Systematic evaluation of proposed tuning algorithm on more human subjects and terrain types (sand, grass, foam, etc.) are needed in order to show the clinical value in the future. In addition, we need more designed experiments to further validate the robustness of policies trained by the proposed approach against realistic conditions of uncertainties or disturbances. Another limitation arose from the feature extraction of the continuous knee profile. We selected four discrete points, as a means of dimension reduction, to characterize knee kinematics in each phase of a single gait cycle. Such a selection dropped the information of kinematics between these points, and thus we had little control over the entire profile except for the feature points. To really reproduce target knee kinematics, we need to explore more advanced feature extraction methods that can better characterize a continuous profile. Lastly, in this study, the impedance tuning goal was to drive the prosthetic knee kinematics to approach the pre-defined knee profile extracted from representative motion in able-bodied population. This is limiting as well as such a goal might not align with user's preference, thus may not be optimal with respect to other measures of gait performance (e.g., gait symmetry, energy expenditure, balance, etc.). How to set up target profiles will be investigated in our future study.
\section{Conclusion}
\label{CONCLD}

In this paper, we proposed an improved solution of an RL-based algorithm, PICE, to learn impedance tuning policies for a robotic knee prosthesis efficiently. The tuning objective was to reproduce near-normal knee kinematics during walking tasks. The PICE algorithm benefited from an ability of offline-online training and from making a compromise in using a simplified value function approximation structure. The problem of falsely yielded negative performance values due to approximation errors was avoided by imposing a PSD constraint to keep the approximated value function qualitatively correct. Therefore, it has great advantage on improving efficiency of the policy training.

We directly tested the proposed idea on human subjects. Our results showed that PICE successfully provided impedance tuning policy to the prosthetic knee with a human in the loop, and it significantly reduced policy training time especially for online training after initializing with an offline pre-trained policy. In addition, the deployed policy is robust across human subjects and modifications in tasks. These promising results suggest great potential for future clinical application of our proposed methods on automatically personalizing assistive wearable robots.

\appendix[Error Bound Analysis]
\label{Proof}
The focus of the analysis is to obtain a qualitative error bound for our proposed PICE algorithm. The analysis is performed along the line of \cite{lagoudakis2003,Kolter2011}. We assume the following sufficient condition that the contraction property holds for any $r\subset\mathbb{R}^m$ under the general projected Bellman operation without constraints
\begin{align*}
  \lVert\text{proj}_{S}(T\bm{\Phi}r)\rVert_\Xi
  \leq\lVert\bm{\Phi}r\rVert_\Xi
  \hspace{1mm} .
\end{align*}
The rationale of the assumption is that the property is theoretically guaranteed by adopting an on-policy scheme, while it can also be practically fulfilled by means of optimized sampling when utilizing an off-policy scheme as demonstrated in \cite{Kolter2011}.

With the above assumption and non-expansive property of projections shown below
\begin{align*}
  \lVert\text{proj}_{S_+}(\cdot)\rVert_\Xi
  \leq\lVert\text{proj}_{S}(\cdot)\rVert_\Xi
  \leq\lVert\cdot\rVert_\Xi
  \hspace{1mm},
\end{align*}
we have the following bounded error between the approximate value function $\hat{Q}^{\pi}$ and the ground truth ${Q}^{\pi}$ for the target policy $\pi$ in the context of PICE
\begin{align*}
  \lVert\hat{Q}^{\pi}-Q^\pi\rVert_\Xi
  \leq&\lVert\hat{Q}^{\pi}-\text{proj}_{S_+}(Q^{\pi})\rVert_\Xi+\lVert\text{proj}_{S_+}(Q^{\pi})-Q^{\pi}\rVert_\Xi\\
  =&\lVert\text{proj}_{S_+}B(\hat{Q}^{\pi})-\text{proj}_{S_+}B(Q^{\pi})\rVert_\Xi\\
  &+\lVert\text{proj}_{S_+}(Q^{\pi})-Q^{\pi}\rVert_\Xi\\\leq&\alpha\lVert\text{proj}_{S}(T\hat{Q}^{\pi})-\text{proj}_{S}(TQ^{\pi})\rVert_\Xi\\
  &+\lVert\text{proj}_{S_+}(Q^{\pi})-Q^{\pi}\rVert_\Xi\\
  \leq&\alpha\lVert\text{proj}_{S}(T\hat{Q}^{\pi})-\text{proj}_{S}(T\text{proj}_{S}(Q^{\pi}))\rVert_\Xi\\
  &+\alpha\lVert\text{proj}_{S}(T\text{proj}_{S}(Q^{\pi}))-\text{proj}_{S}(TQ^{\pi})\rVert_\Xi\\
  &+\lVert\text{proj}_{S_+}(Q^{\pi})-Q^\pi\rVert_\Xi\\\leq&\alpha\lVert\hat{Q}^{\pi}-{Q}^{\pi}\rVert_\Xi+\alpha\kappa(\Theta)\lVert\text{proj}_{S}(Q^{\pi})-Q^{\pi}\rVert_\Xi\\
  &+\lVert\text{proj}_{S_+}(Q^{\pi})-Q^\pi\rVert_\Xi\\
  \leq&\frac{1+2\alpha\kappa(\Theta)}{1-\alpha}\lVert\text{proj}_{S_+}(Q^{\pi})-Q^\pi\rVert_\Xi
  \hspace{1mm},
\end{align*}
where the $\kappa(\Theta)$ denotes the condition number of the matrix $\Theta\triangleq(\Xi\Lambda)^{-\frac{1}{2}}$. The $\Xi$ and $\Lambda$ are diagonal matrices consisting of steady-state probabilities of Markov chains under the behavior and target policy, respectively. 

Let $\xi$ be the largest upper bound of evaluation errors over all iterations of policy evaluation. According to \cite{bertsekas2012Book,lagoudakis2003}, when the policy improvement is performed exactly without incurring errors (which is guaranteed in this problem setting due to a use of quadratic programming solution), the following bound is yielded
\begin{equation*}
\limsup_{i\to\infty}\parallel\hat{Q}^{(i)}-Q^{\ast} \parallel\leq\frac{2\alpha\xi}{(1-\alpha)^2} \hspace{1mm}.
\end{equation*}

This result for off-policy learning may result in a less tight bound if the condition number is poor. However, it still provides performance guarantees for the solution quality of PICE with the error being bounded, as opposed to being arbitrarily large. The bound implies that the proposed algorithm will eventually either converge or oscillate within a suboptimal policy space where the resulted policy is at most a constant away from the true optimality.

\bibliographystyle{IEEEtran}
\bibliography{IEEEabrv,biblist.bib}

\ifCLASSOPTIONcaptionsoff
  \newpage
\fi

\end{document}